\documentclass[twoside,slac_one]{revtex4}
\usepackage{graphicx}
\usepackage{fancyhdr}
\usepackage{amsmath} 
\usepackage{bm}
\usepackage{amsxtra}
\usepackage{amssymb}
\usepackage{amsthm}
\usepackage{latexsym}
\usepackage{lscape}
\usepackage{placeins}

\usepackage{xspace}

\def\NIMA{{\em Nucl.\,Instrum.\,Methods}\,A}

\def\PLB{{\em Phys.\,Lett.}\,B}
\def\PRL{{\em Phys.\,Rev.\,Lett.}}
\def\PRD{{\em Phys.\,Rev.}\,D}

\def\JINST{{\em JINST}}
\def\JPG{{\em J.\,Phys.}\,G}

\newcommand{\decay}[2]{\ensuremath{#1\!\to #2}\xspace}         

\def\BuToJPsiK {\decay{\Bu}{\jpsi\Kp}\xspace}

\newcommand{\CL}{C.L.\ }

\newcommand{\CLs}{\ensuremath{\textrm{CL}_{\textrm{s}}}\xspace}
\newcommand{\CLb}{\ensuremath{\textrm{CL}_{\textrm{b}}}\xspace}

\newcommand{\bbdim}{\ensuremath{b\bar{b}\to \mu \mu X}\xspace}

\newcommand{\Bq}{\ensuremath{B^0_q}\xspace}
\newcommand{\Bs}{\ensuremath{B^0_s}\xspace}
\newcommand{\Bd}{\ensuremath{B^0}\xspace}
\newcommand{\Bu}{\ensuremath{B^+}\xspace}

\newcommand{\Bsmumu}{\ensuremath{\Bs\to\mu^+\mu^-}\xspace}
\newcommand{\Bdmumu}{\ensuremath{\Bd\to\mu^+\mu^-}\xspace}
\newcommand{\Bmumu}{\ensuremath{B^0\to\mu^+\mu^-}\xspace}
\newcommand{\Bdsmumu}{\ensuremath{B^0_{d,s}\to\mu^+\mu^-}\xspace}

\newcommand{\BsKK}{\ensuremath{\Bs\to K^+K^-}\xspace}

\newcommand{\BdKpi}{\ensuremath{\Bd\to K^+\pi^-}\xspace}

\newcommand{\Bhh}{\ensuremath{B^0_{q}\to h^+h^{'-}}\xspace}
\newcommand{\Bmm}{\ensuremath{B^0_{(s)}\to \mu^+\mu^-}\xspace}

\newcommand{\BuJpsiK}{\ensuremath{B^+\to J/\psi K^+}\xspace}

\newcommand{\Jpsi}{\ensuremath{J\!/\!\psi}\xspace}
\newcommand{\jpsi}{\Jpsi}
\def\BsToJPsiPhi  {\decay{\Bs}{\jpsi\phi}\xspace}

\def\BdToKpi      {\decay{\Bd}{\Kp\pi^-}\xspace}
\def\Kp    {\ensuremath{K^+}\xspace}

\newcommand{\Bqmumu}{\ensuremath{\ensuremath{B^0_{q}}\to\mu^+\mu^-}\xspace}

\newcommand{\BsJpsiPhi}{\ensuremath{B^0_s\to J/\psi \phi}\xspace}

\newcommand{\BRof}[1]{\ensuremath{{\cal B}(#1)}\xspace}

\newcommand{\mevc}{\ensuremath{\,{\rm MeV}\!/\!c}\xspace}
\newcommand{\gevc}{\ensuremath{\,{\rm GeV}\!/\!c}\xspace}
\newcommand{\mevcc}{\ensuremath{\,{\rm MeV}\!/\!c^2}\xspace}
\newcommand{\gevcc}{\ensuremath{\,{\rm GeV}\!/\!c^2}\xspace}

\newcommand{\invpb}{\ensuremath{\,{\rm pb}^{-1}}\xspace}

\begin{document}

\title{\boldmath Search for \Bdsmumu at LHCb with 300\,pb$^{-1}$}

%

\author{M.-O. Bettler}
\affiliation{Istituto Nazionale di Fisica Nucleare (INFN), Sezione Firenze, Italy}

\begin{abstract}
A search for the  \Bsmumu and \Bdmumu decays is performed in $\sim 300$\,pb$^{-1}$ of $pp$ collisions at $\sqrt{s}$ = 7\,TeV
collected by the LHCb experiment at the Large Hadron Collider at CERN.  
The measured limit on the branching fraction of the \Bsmumu decay is \BRof \Bsmumu $< 1.6 \times 10^{-8}$ at 95\,\% confidence level, while that of the \Bdmumu decay  is \BRof\Bdmumu $< 5.1 \times 10^{-9}$ at 95\,\% confidence level.\\
A combination with the 2010 dataset of LHCb yields a limit of \BRof\Bsmumu $<1.5 \times 10^{-8}$ at 95\,\% confidence level.
\end{abstract}

\maketitle

\thispagestyle{fancy}

\section{Introduction}\label{sec:introduction}

Measurements at low energies may provide interesting indirect constraints on the masses of particles that are too heavy to be produced directly.
This is particularly true for Flavor Changing Neutral Currents (FCNC) processes which are highly suppressed in the Standard Model (SM) and can only occur through higher order diagrams.
 
The SM prediction for the branching fractions ($\mathcal{B}$) of the FCNC decays \Bsmumu and \Bdmumu~\footnote{In these proceedings the inclusion of charge-conjugate states is implicit.} have been computed~\cite{Buras2010}  to be \BRof \Bsmumu = $(3.2 \pm 0.2) \times 10^{-9}$ and \BRof \Bdmumu = $(0.10 \pm 0.01) \times 10^{-9}$.
However New Physics (NP) contributions can significantly enhance these values.
For instance, within Minimal Supersymmetric extensions of the SM (MSSM), in the large $\tan \beta$ approximation~\cite{MSSM}, \BRof \Bsmumu is found to be proportional to $\sim \tan^6\beta$, where $\tan\beta$ is the ratio of the vacuum expectation values of the two neutral CP-even Higgs fields.
Therefore it could be strongly enhanced for large values of $\tan \beta$.

The best published limits from the Tevatron at $95\%$\CL  are obtained using 6.1 fb$^{-1}$ by the D0 collaboration~\cite{d0_PLB}, \BRof \Bsmumu $<5.1 \times 10^{-8}$ and using 2 fb$^{-1}$ by the CDF collaboration~\cite{cdf_PRL}, \BRof \Bsmumu $<5.8 \times 10^{-8}$ and \BRof \Bdmumu $<1.8 \times 10^{-8}$.
The CDF collaboration has also presented a preliminary result~\cite{cdf_bsmumu_preprint} with 6.9 fb$^{-1}$ that lowers the limit  to \BRof \Bdmumu$<6.0 \times 10^{-9}$ at 95\% \CL. 
In the same dataset an excess of \Bsmumu candidates is also observed compatible with a \BRof \Bsmumu = $(1.8^{+1.1}_{-0.9}) \times 10^{-8}$ and with an upper limit \BRof \Bsmumu $< 4.0 \times 10^{-8}$ at 95\,\% C.L.
The CMS collaboration has recently presented a preliminary result~\cite{cms_preprint} with 1.14\,fb$^{-1}$ and set the limits  \BRof \Bsmumu $< 1.9 \times 10^{-8}$ and  \BRof \Bdmumu $< 4.6 \times 10^{-9}$ at 95\,\% C.L.

The LHCb collaboration has obtained limits \cite{LHCb_paper} comparable with the Tevatron ones, \BRof \Bsmumu $<5.4 \times 10^{-8}$ and  \BRof \Bdmumu$<1.5 \times 10^{-8}$ at 95$\%$ \CL based on 37 pb$^{-1}$ of luminosity collected in the 2010 run.
This document presents an analysis of the LHCb data taken with an integrated luminosity of 300 pb$^{-1}$ in 2011. A combination of this data with the data used for the analysis published on the 2010 data based on 37 pb$^{-1}$ is also performed.

\section{The LHCb detector}
\label{sec:detector}

The LHCb detector~\cite{LHCbdetector} is a single-arm forward spectrometer with an acceptance for charged tracks with pseudorapidity $\eta$ of $2 < \eta < 5$.
It consists of a vertex locator, a warm dipole magnet, a tracking system, two RICH detectors, a calorimeter system and a muon system. 

Track momenta are measured to a precision of $\delta p / p$ ranging  $0.35$--$0.5\,\%$. 
The RICH system provides charged hadron identification in a momentum range 2--100\gevc.
Typically, kaon identification efficiencies of over 90\% can be attained for a $\pi \to K$ fake rate below 10\%.
The calorimeter system identifies high transverse energy ($E_{\rm T}$) hadron, electron and photon candidates and provides information for the trigger.
The muon system provides information for the trigger and muon identification with an efficiency of $\sim 95$\,\% for a misidentification rate of about 1--2\,\% for momenta above 10\gevc.

\section{Analysis Strategy}

The analysis presented in this document uses about 300 pb$^{-1}$ of data collected by LHCb between March and June 2011 at $\sqrt{s}$ = 7\,TeV. 
Assuming the SM branching fractions and  $\sigma_{b \overline{b}} = 75\pm14\,\mu$b as measured within the LHCb acceptance~\cite{bbxsection}, approximately $3.4~(0.32)$ 
$B^0_{s}\to \mu^+\mu^-$ ($B^0 \to \mu^+\mu^-$)  events are expected to be reconstructed and selected in this sample.
The general structure of the analysis is unchanged with respect to Ref.~\cite{LHCb_paper}, but has been improved in two main points: the optimization of the multi-variate analysis discriminant using geometrical and kinematic variables and the use of the RICH particle identification (PID) in the analysis.

The selection procedure treats signal and control/normalization channels in the same way in order to minimize systematic uncertainties.
After the selection, each event is given a probability to be signal or background in a two-dimensional space defined by the invariant mass and a Boosted Decision Tree (BDT) from the TMVA package~\cite{tmva}.
The combination of variables entering the BDT is optimized using Monte Carlo (MC) simulation. 
However the probability for a signal or background event to have a given value of the BDT is obtained from data using trigger unbiased \Bhh events for signal and sideband \Bmm candidates for background.

The parameters describing the invariant mass line shape  of the signal are extracted from data using control samples.
The average mass values are obtained from \BdKpi and \BsKK exclusive samples. 
The \Bs and \Bd mass resolutions are estimated by interpolating the ones obtained with the dimuon resonances ($J/\psi, \psi(2S)$ and $\Upsilon(1S),\Upsilon(2S)$ and $\Upsilon(3S)$) and cross-checked via a fit to the invariant mass distribution of the \Bhh inclusive decays and of the \BdKpi exclusive decay. 

The number of expected signal events is obtained by normalizing to channels of known branching fractions, \BuToJPsiK, \BsToJPsiPhi and \BdToKpi that are selected in a way as similar as possible to the signals in order to minimize the systematic uncertainty due to the different phase space.
The probability for a background event to have a given BDT and invariant mass value is obtained by a fit of the mass distribution of events in the mass sidebands, in bins of BDT.
The two-dimensional space formed by the invariant mass  and BDT is binned, and for each bin we compute how many events are observed in data, how many signal events are expected for a given BF hypothesis and luminosity, and how many background events are expected for a given luminosity.
The compatibility of the observed distribution of events in all bins with the expected one for a given BF hypothesis is computed using the modified frequentist method method~\cite{Read_02}, which allows to exclude a given hypothesis at a given confidence level.
In order to avoid unconscious biases, the data in the mass region defined by $M_{\Bd} - 60 \mevcc $ and  $ M_{\Bs} + 60 \mevcc$ have been blinded until the completion of the analysis.

\section{Event selection}

The selection aims at reducing the data size to a manageable level while keeping the efficiency on the signal as high as possible.
The selection consists of loose requirements on track separation from the interaction point, decay vertex quality, compatibility of the reconstructed origin of the $B$ meson with the interaction point and muon identification. 
Events passing the selection are considered  $B^0_{(s)} \to \mu^+ \mu^-$ candidates if their invariant mass lies  within 60\mevcc of the nominal $B^0_{(s)}$ mass.

A similar selection is applied to the normalization channels, in order to minimize systematic errors in the ratio of efficiencies.
The \Bhh inclusive sample is selected exactly in the same way as the \Bmm signals (apart from the muon identification requirement) in order to be used as the main control sample in the calibration of the BDT and the invariant mass. 
The $J/\psi \to \mu \mu$ decay in the \BuJpsiK and \BsJpsiPhi normalization channels is selected 
exactly in the same way as \Bmm signals.
$K^{\pm}$  candidates are required to pass some track quality and impact parameter cuts.

Fiducial cuts are applied to reject fake signal candidates. 
All tracks from selected candidates are required to have a momentum less than 1\,TeV/$c$. 
Furthermore, only $B$ candidates with lifetimes less than $5 \times \tau_{B_{(s)}}$ are accepted for further analysis.
Dimuon candidates coming from elastic di-photon production are effectively removed by  requiring  a minimal transverse momentum of the $B$ candidate of 500\mevc.

\section{Signal and background likelihoods}
\label{sec:likelihoods}

The discrimination of the signal from the background is achieved through the combination of two independent variables: the BDT and the invariant mass. 
The invariant mass in the search regions ($\pm 60\mevcc$ around the $B^0_{(s)}$  masses) is divided into six equal-width bins, and the BDT into four equal-width bins distributed between zero and one. 

The BDT combines information related with the topology and kinematics of the event: the $B^0_{(s)}$ lifetime, the minimum impact parameter of the two muons, the distance of closest approach of the two tracks, the $B^0_{(s)}$ impact parameter  and $p_T$ and the isolation of the muons with respect to the other tracks of the event, the cosine of the polarization angle, the $B$ isolation, the minimum $p_{\rm T}$ of the two muons.
Those variables have been chosen on the basis of the performance on MC samples.
The BDT output is found to be fully independent of the invariant mass both for signal and for background.
The BDT is trained using Monte Carlo simulations (\Bmm for signals and \bbdim for background) and calibrated with data using $B^0_{(s)}\to h^+h^{'-}$ selected as the signal events and triggered independently on the signal in order to avoid trigger bias.

\subsection{Calibration of the invariant mass likelihood for signal}

The invariant mass shape for the signal is parametrized as a Crystal Ball function. 
The mass average value and the mass resolution are calibrated from data and the transition point of the radiative tail is taken from Monte Carlo simulation.

\begin{figure}[t]
\centering
\includegraphics[width=.48\textwidth]{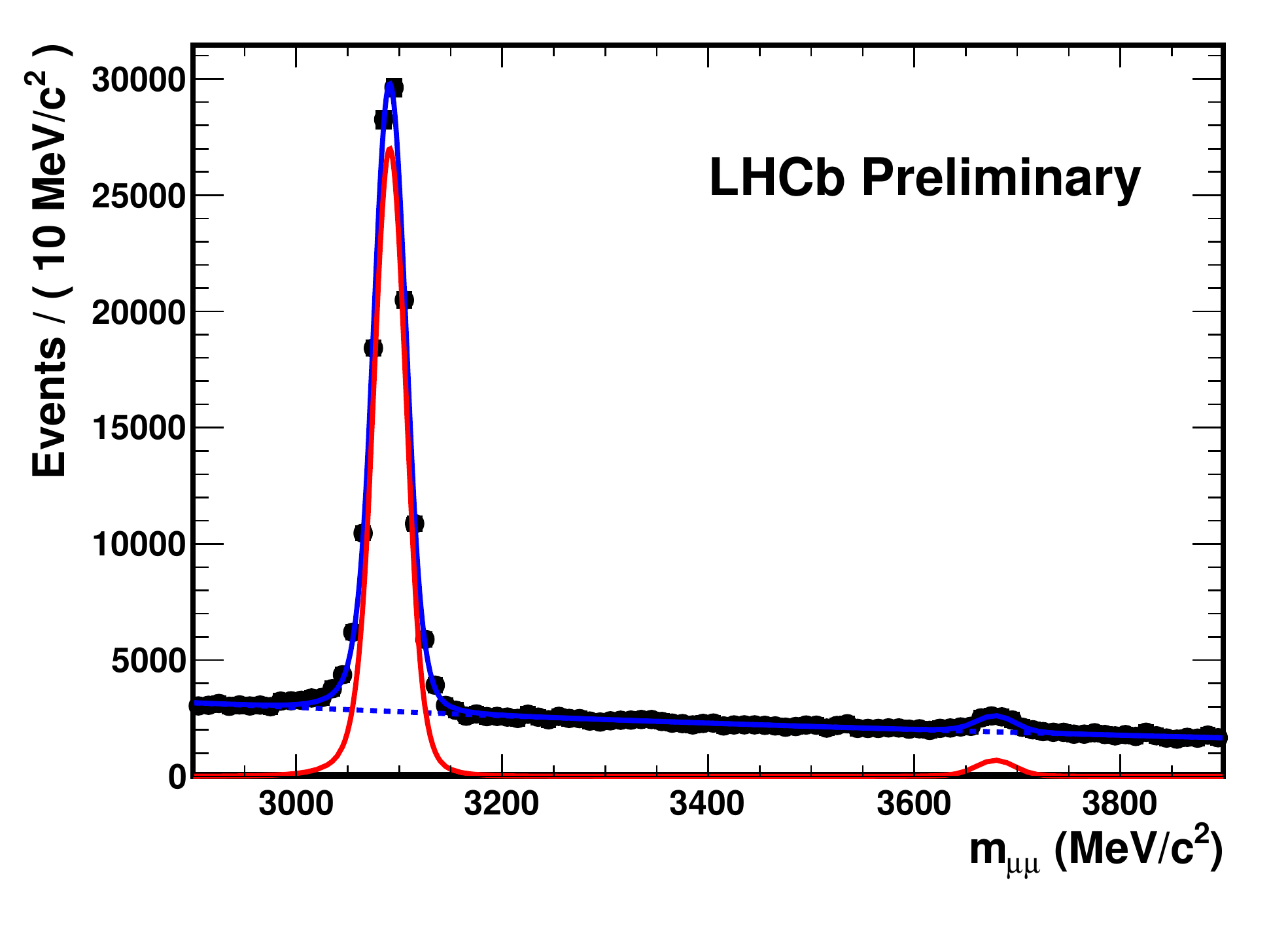}
\includegraphics[width=.48\textwidth]{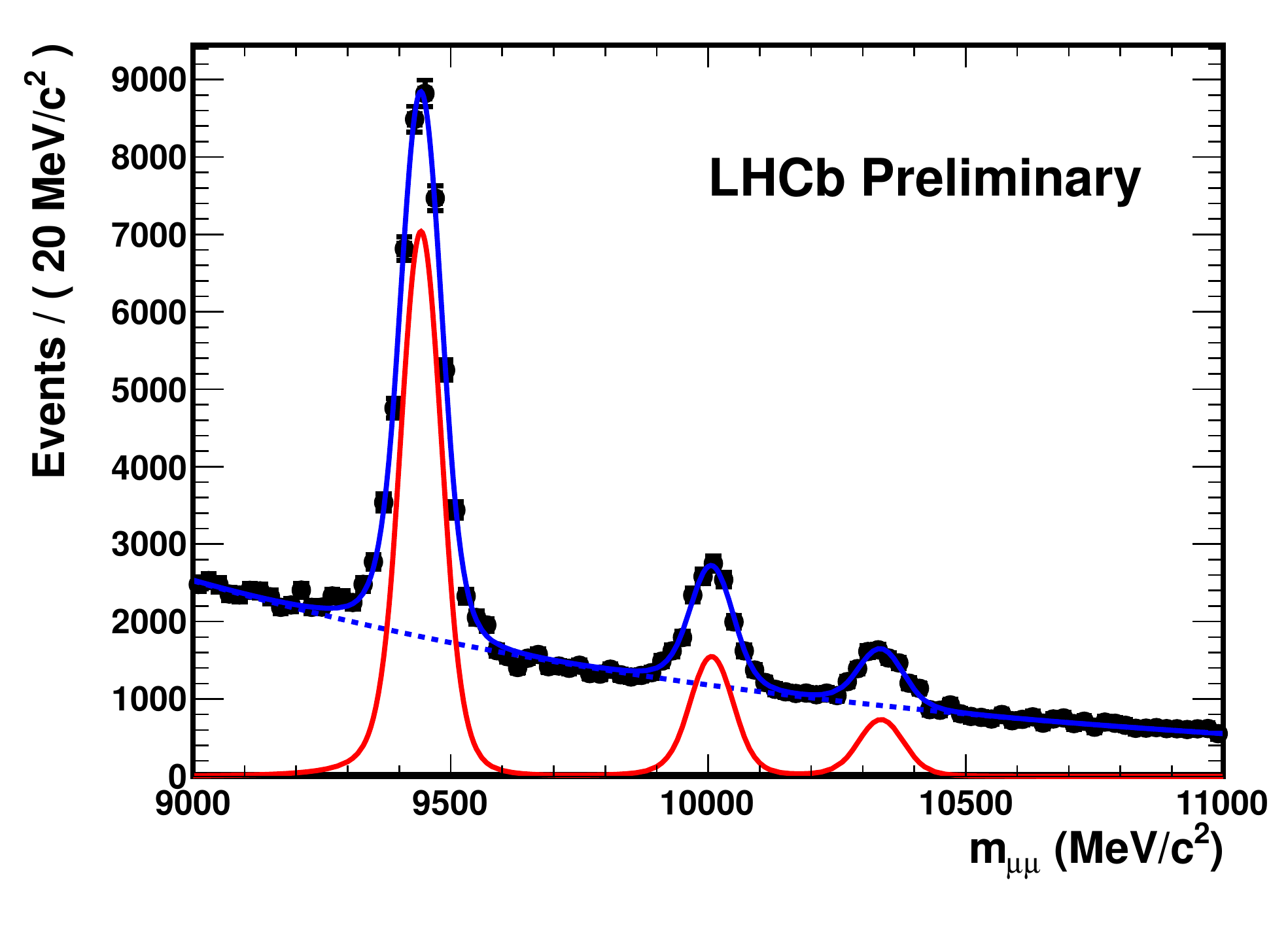}
\vspace{-4mm}
\caption{Fits to the dimuon invariant mass spectrum in the $[2.9, 3.9]\gevcc$ range (left) and in $[9,11]\gevcc$ (right).}
\label{fig:dimuon_reso}
\end{figure}

The average mass values are extracted from \BdKpi and \BsKK exclusive candidates
and the results are the following:
\begin{eqnarray}
 m(\Bs) & = & (5358.0 \pm 1.0) \mevcc\nonumber \\ 
 m(\Bd) & = & (5272.0 \pm 1.0) \mevcc. \nonumber
\end{eqnarray}
The mass resolutions are extracted from data by a linear interpolation between the measured resolution of charmonium and bottomonium resonances decaying to two muons (\jpsi, $\psi$(2S) and $\Upsilon(1S)$, $\Upsilon(2S)$, $\Upsilon(3S)$) as shown in Fig.~\ref{fig:dimuon_reso}.
This result has been cross-checked with the mass resolution obtained via full fit to the \Bhh inclusive decay and the \BdKpi exclusive decay. 
The interpolation yields $\sigma(\Bs) =  (24.6 \pm 0.2_{\rm stat} \pm 1.0_{\rm syst}) \mevcc$.

\subsection{Calibration of the BDT likelihood for signal}

\begin{figure}[t]
\centering
\includegraphics[width=.42\textwidth]{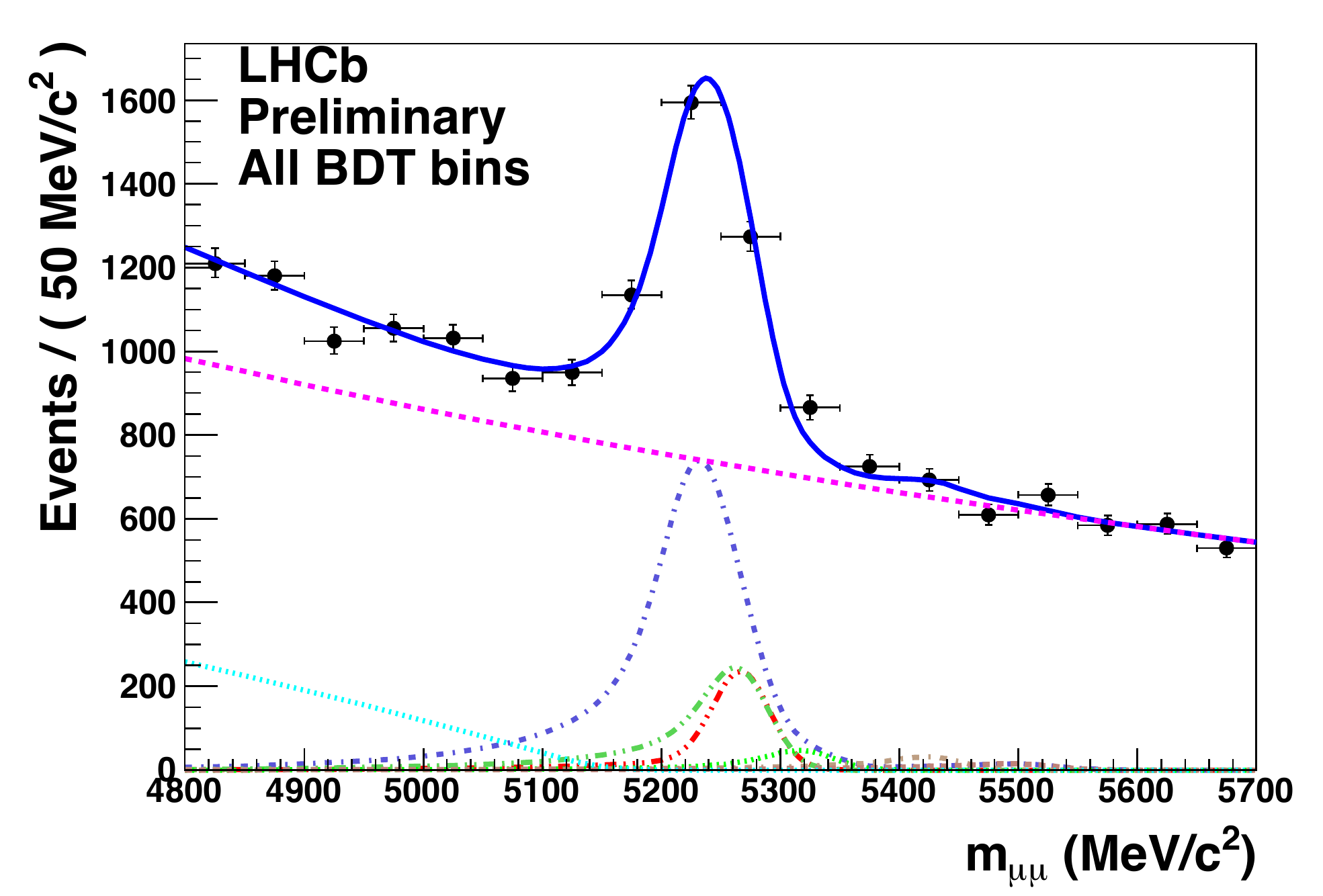}
\includegraphics[width=.42\textwidth]{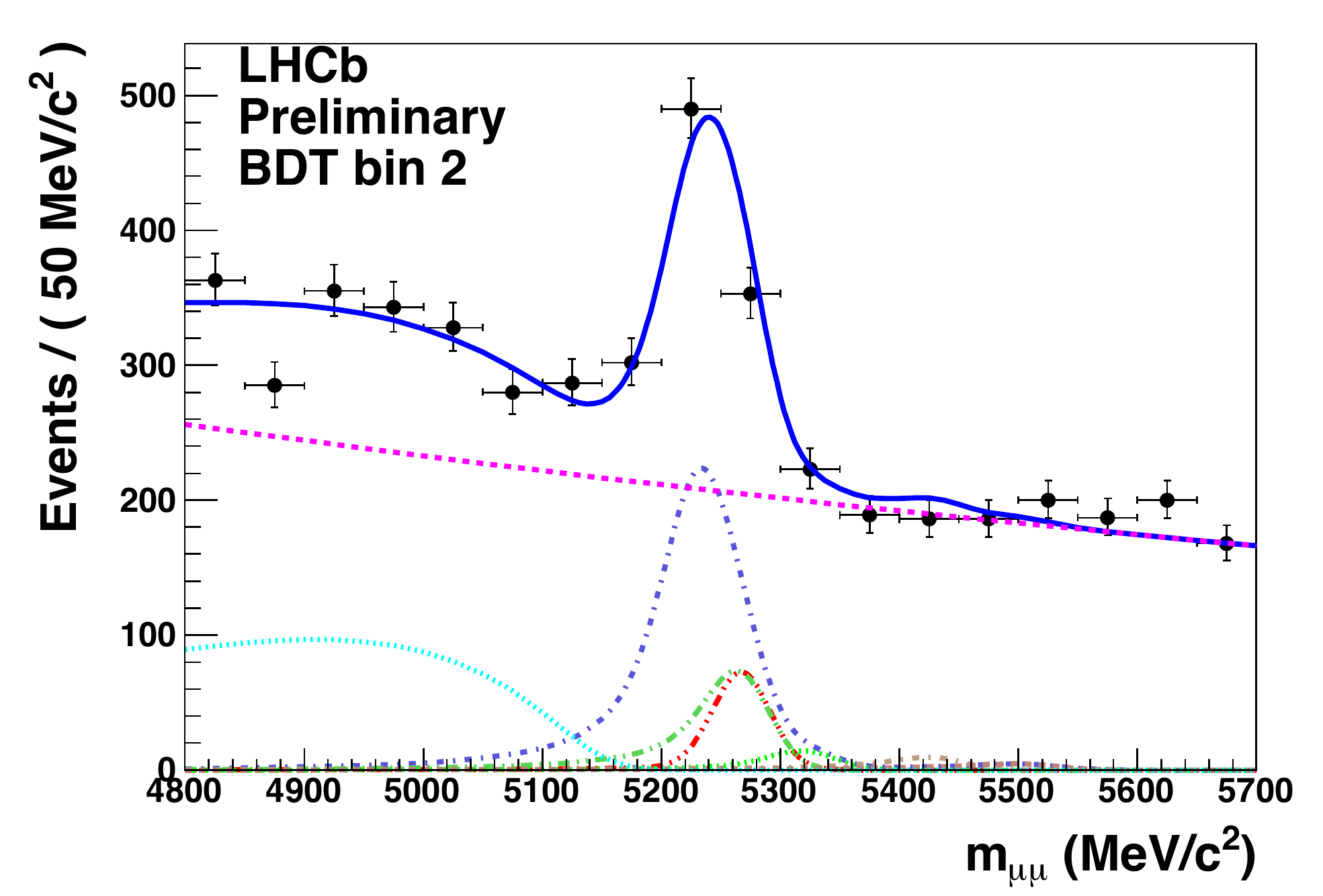}
\includegraphics[width=.42\textwidth]{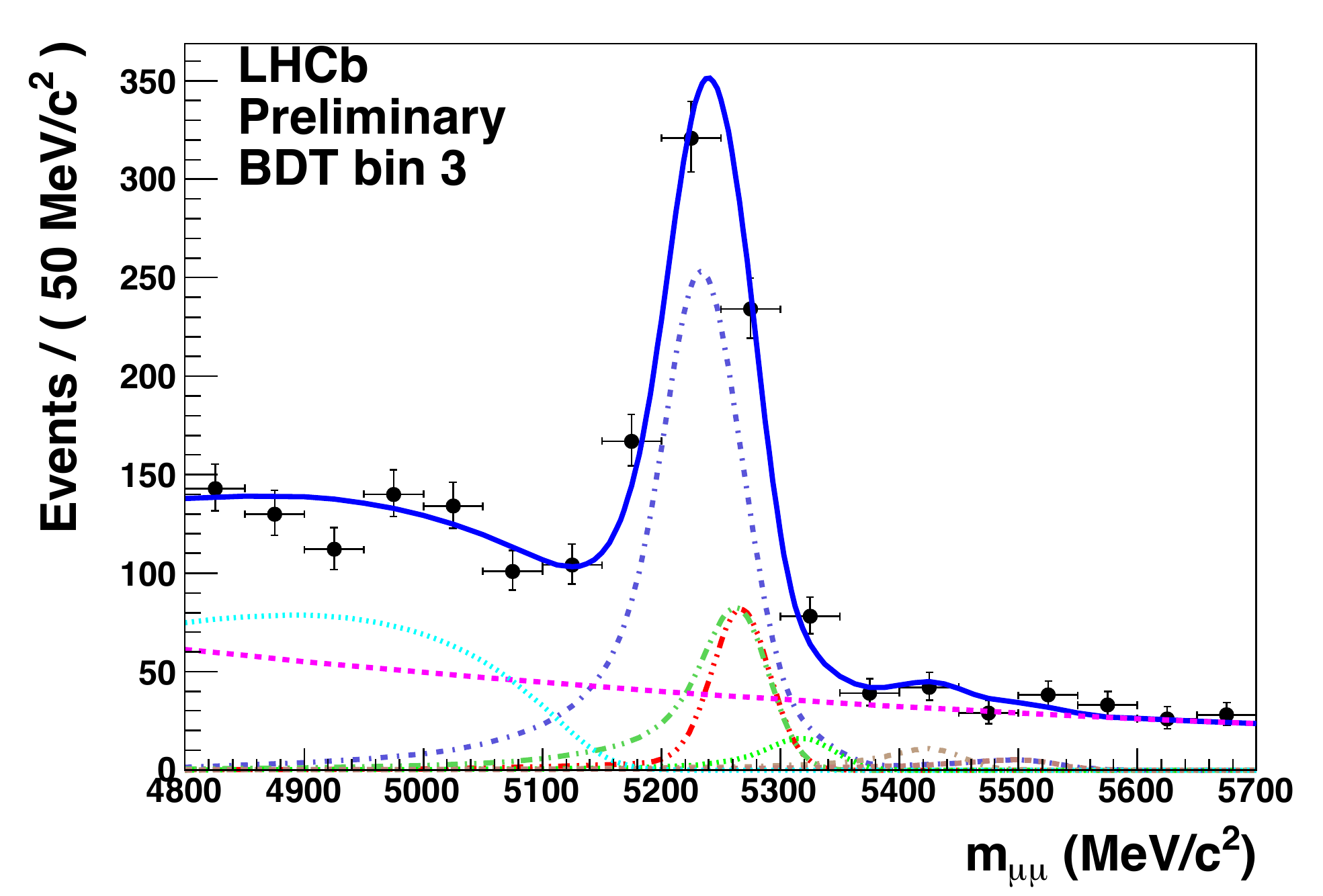}
\includegraphics[width=.42\textwidth]{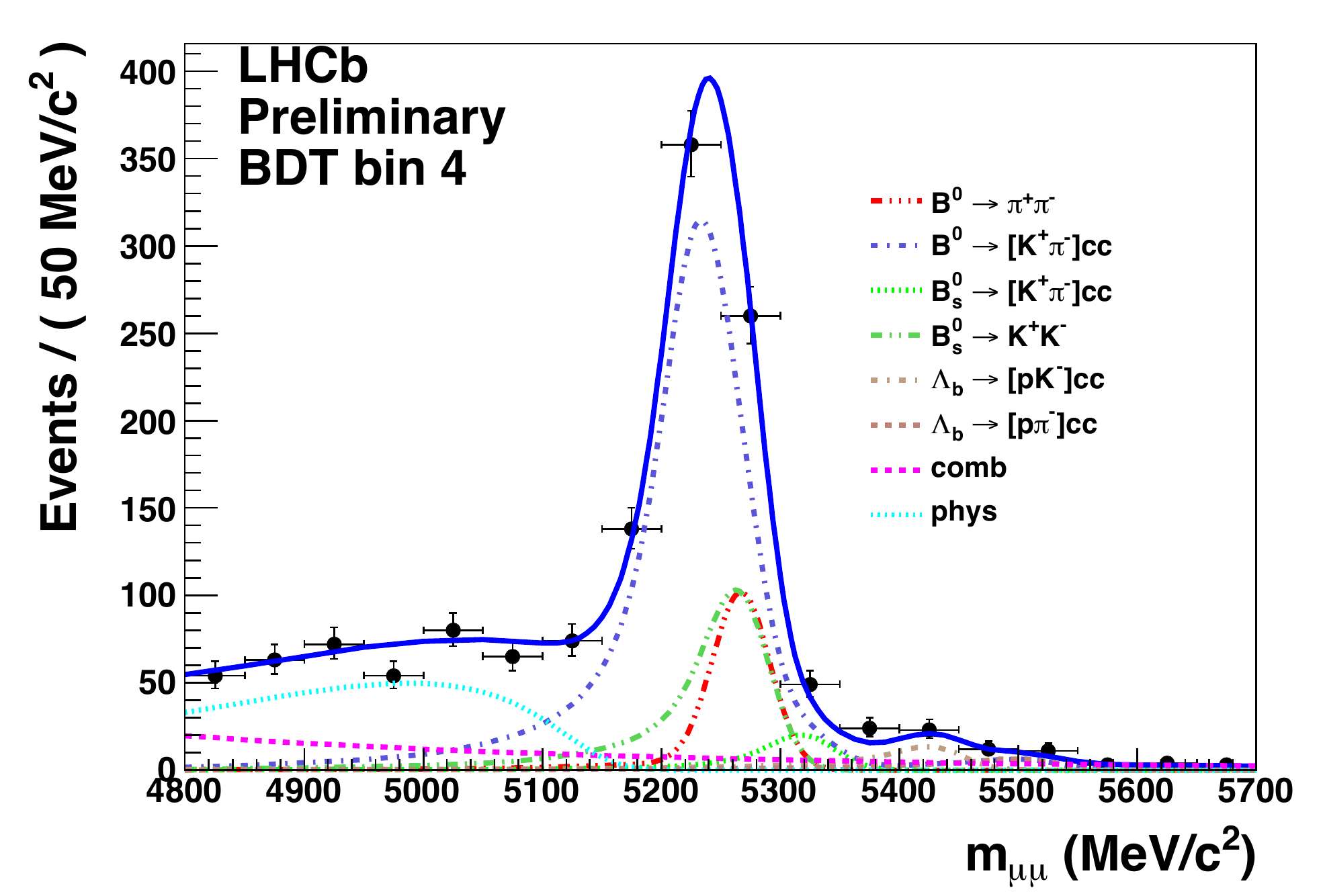}
\vspace{-4mm}
\caption{Full fit to the invariant mass distributions of \Bhh candidates 
in the $\pi\pi$ mass hypothesis for the whole sample (top left) and for the samples
in bins 2,3,4 of the BDT (top right, bottom left, bottom right).}
\label{fig:b2hh_fitinbins}
\end{figure}

\begin{figure}[t]
\centering
\includegraphics[width=.45\textwidth]{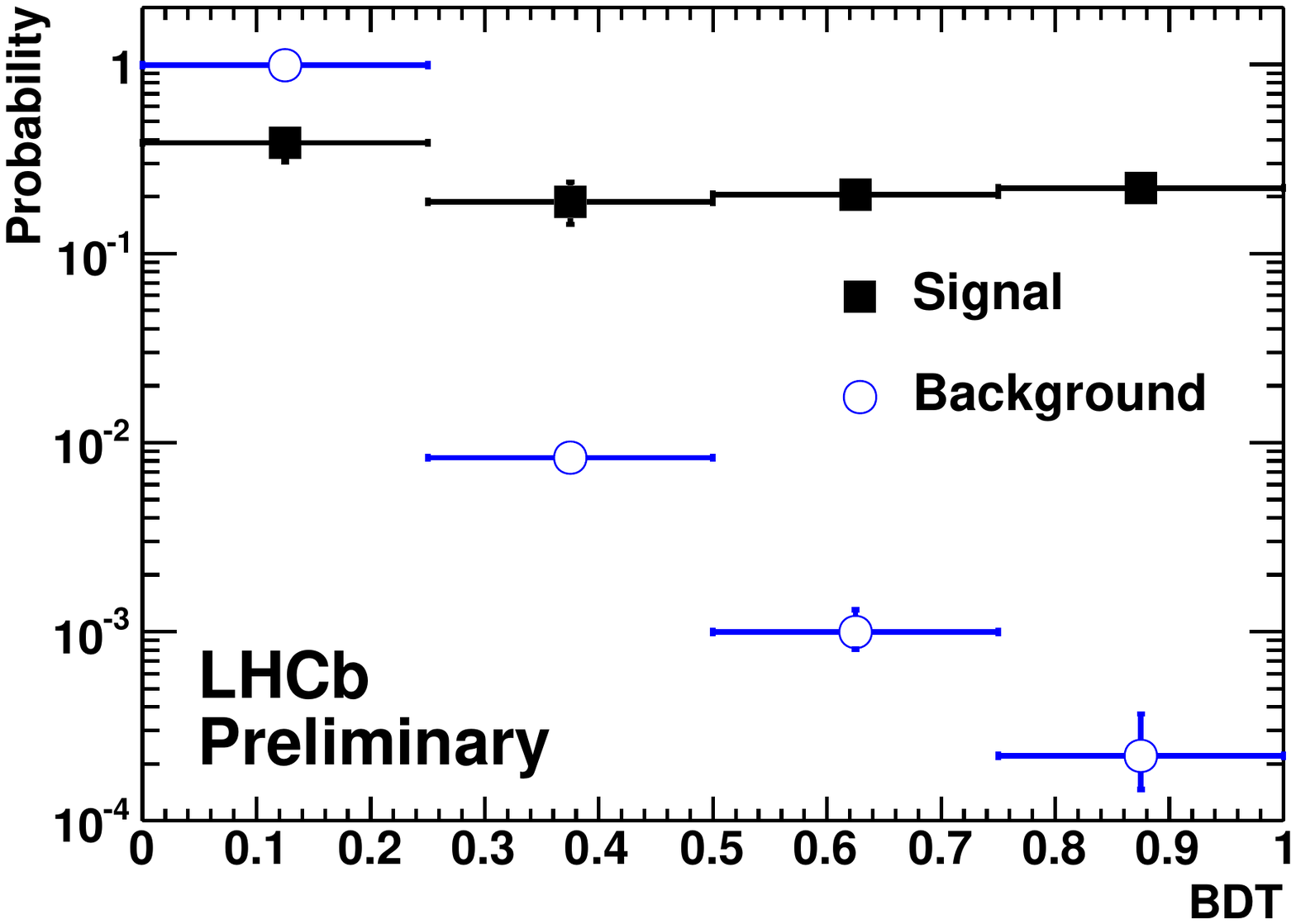}
\vspace{-4mm}
 \caption{Probability of signal events in bins of BDT obtained from the inclusive sample of TIS \Bhh events (solid squares). 
The background probability (open circles) is obtained from the events in the sidebands.} 
\label{fig:BDT_all}
\end{figure}

The BDT shape for signal events is calibrated using a \Bhh inclusive sample composed only  of events {\it Triggered Independently of the Signal} (TIS)\footnote{Here we consider events that are TIS with respect to the two first trigger levels (L0 and HLT1) but accept all HLT2 triggers. This is due to the lack of statistics for full (L0 $\times$ HLT1 $\times$ HLT2) TIS events. Corrections are computed from Monte Carlo simulations and cross-checked on control channels.} to avoid biases introduced by the hadronic trigger.
The number of \Bhh signal events in each BDT bin is extracted by fitting the $hh$ invariant mass distribution in the $\pi\pi$ mass hypothesis~\cite{Bhhnote}.
Figure~\ref{fig:b2hh_fitinbins} depicts fits to the mass distribution of the full sample and for the three highest BDT bins for \Bhh TIS events is shown.
Those yields are used to compute fractional yield of events as a function of the BDT bin as shown on Fig.~\ref{fig:BDT_all}.
Cross-checks consist of one-dimensional fits to the inclusive sample and of fits to   $B^0_{(s)} \to KK,K\pi, \pi\pi$ exclusive samples selected using the $K - \pi$ separation capability of the RICH system.
The maximum spread among the different methods has been used as a systematic error in the fractional yield determination.

\subsection{Calibration of the combinatorial background Likelihood}

The BDT and invariant mass shapes for the combinatorial background inside the signal regions are calibrated with data by interpolating the number of expected events using fit to the invariant mass sidebands in each bin of BDT.
The lower sideband is potentially polluted by two contributions other than combinatorial background that can affect the fit:  cascading $b \to c \mu \to \mu \mu X$ decays below 4900\mevcc and \Bhh events with the two hadrons misidentified as muons above 5000\mevcc. 
Thus the model used to extract the expected number of background events is chosen to be a single exponential fitted to the events in the region [4900, 5000]\mevcc and in the upper sideband [5418, 6000]\mevcc.

\begin{figure}[t]
\centering
\includegraphics[width=.42\textwidth]{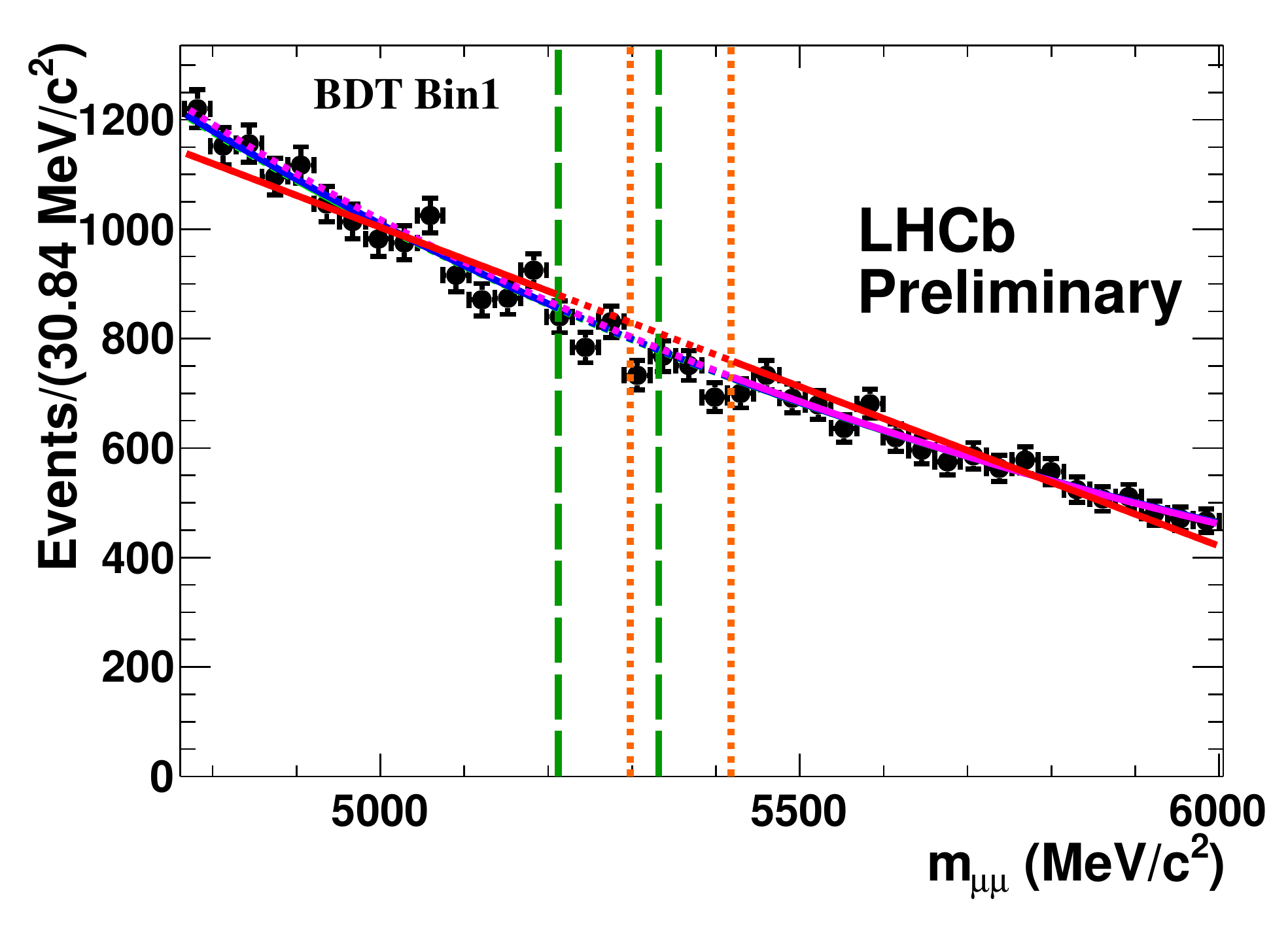}
\includegraphics[width=.42\textwidth]{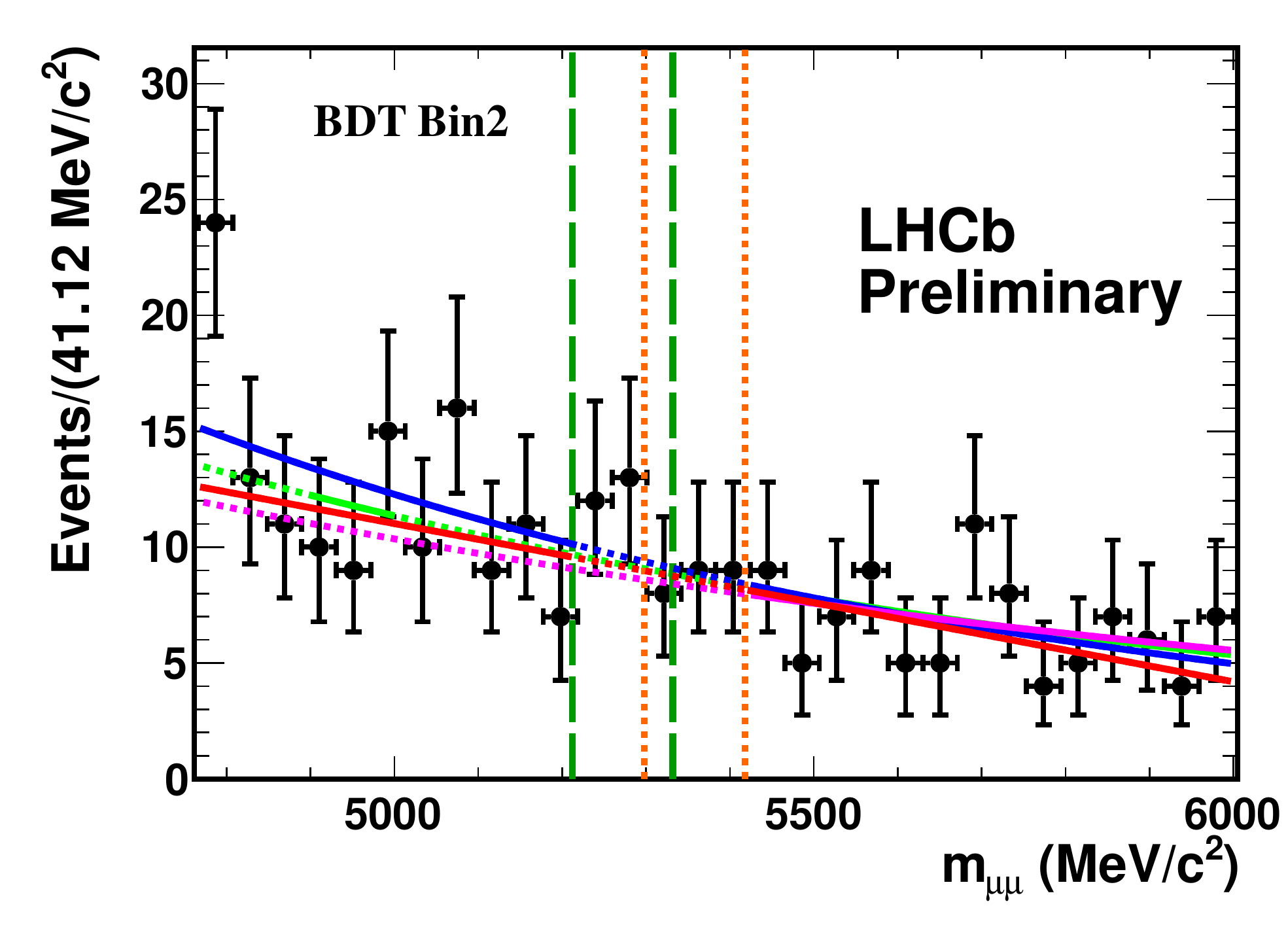}
\includegraphics[width=.42\textwidth]{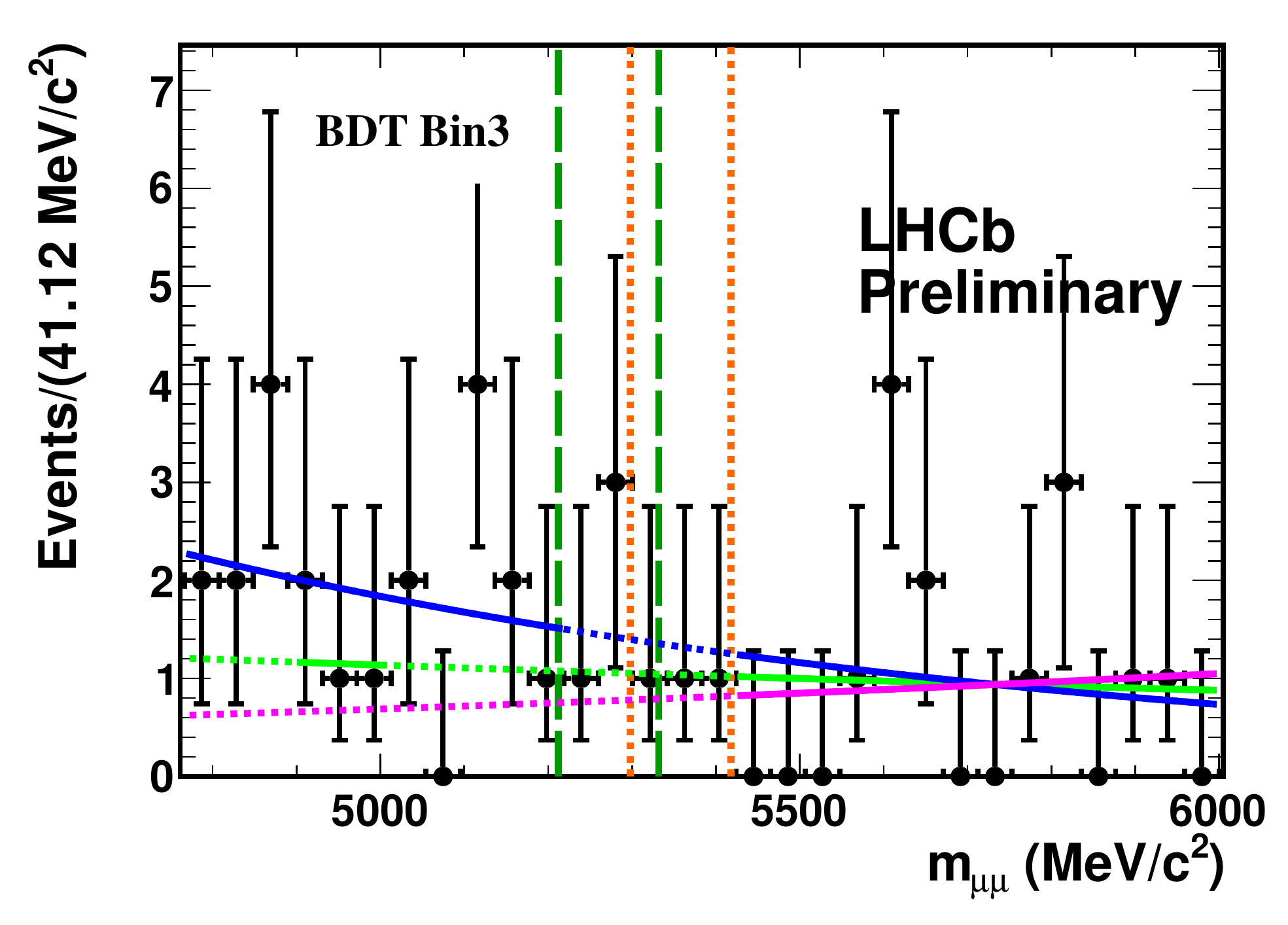}
\includegraphics[width=.42\textwidth]{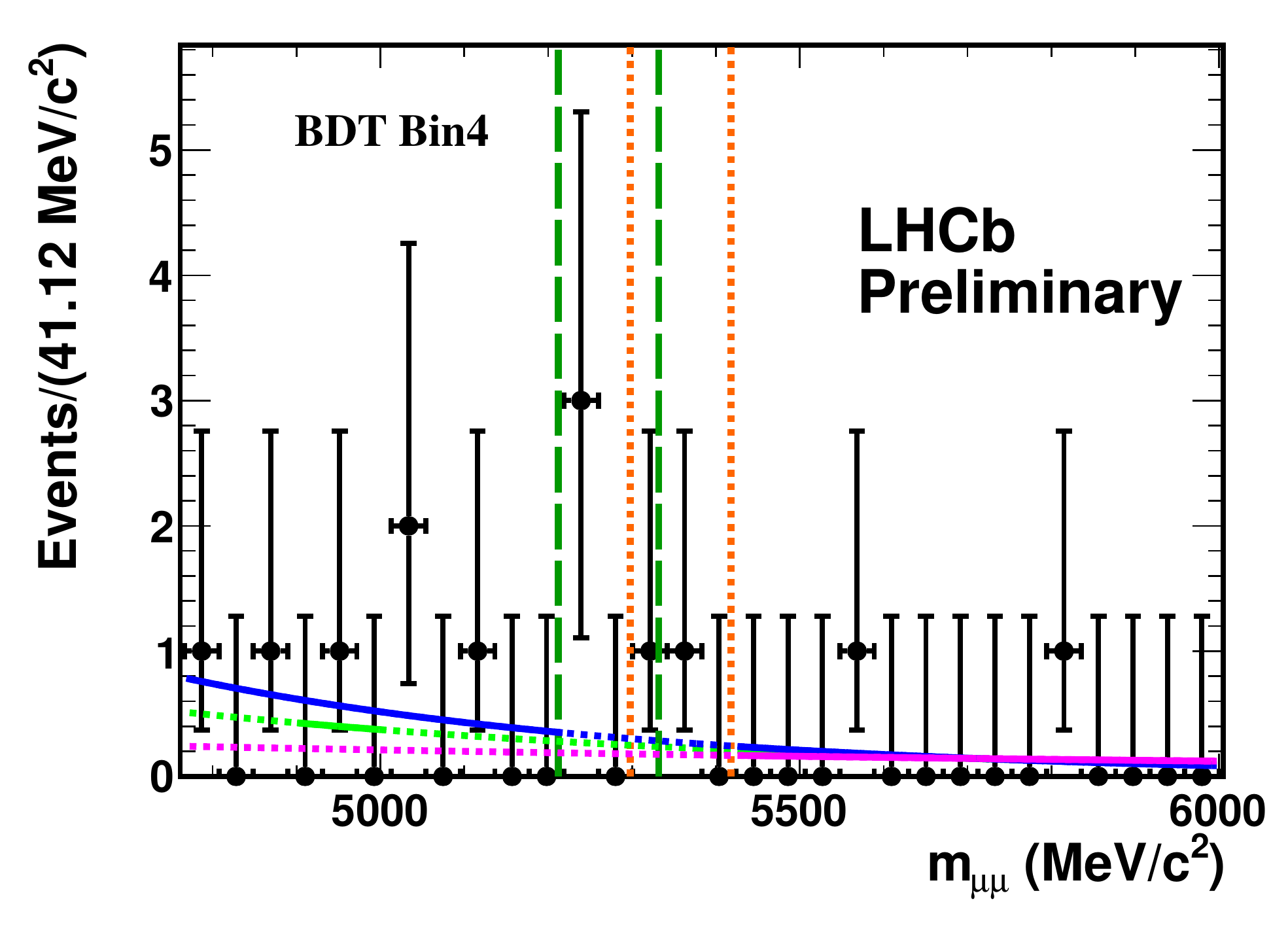}
\vspace{-4mm}
\caption{
Distribution of the $\mu\mu$ invariant mass for events in each BDT bin.
The curves show the models used to fit the sidebands and extract the expected number 
of background events in the signal regions.
Red: linear fit in [4672,5212] \mevcc and [5418,6000]\mevcc sidebands.
Blue: single exponential fit in the same ranges as the red curve.
Green: single exponential fit in the region [4900,5000] \mevcc and [5418,6000]\mevcc.
Magenta: single exponential fit using only the upper sideband ([5418,6000]\mevcc).
The green curve defines the model and ranges used in the analysis.
The  \Bs and \Bd search windows, delimited by the vertical green and orange lines, 
have not been considered in the fitting procedure.}
\label{fig:BDT_bkg_bins}
\end{figure}

In Fig.~\ref{fig:BDT_bkg_bins} the invariant mass distribution for events in each BDT bin is shown and  different fit models used for estimating the expected number of background events in the signal regions are superimposed. 
The fraction of combinatorial background expected per BDT bins is depicts on  Fig.~\ref{fig:BDT_all}.


\subsection{Peaking background Likelihood}

The peaking background coming from \Bhh events for which the two hadrons in the final states are misidentified as muons has been evaluated from data to be $N_{\Bs} = (0.5\pm 0.4)$ events in the \Bs mass window and $N_{\Bd} = (2.5 \pm 2.0)$ events in the \Bd one in the whole BDT range.
The mass lineshape of the peaking background is obtained from a MC sample of double misidentified \Bhh events and normalized to the number of events expected in the two search windows from data.

\section{Normalization}
\label{sec:norm_factors}

\begin{figure}[t] 
\begin{center}
 \includegraphics*[width=0.44\textwidth]{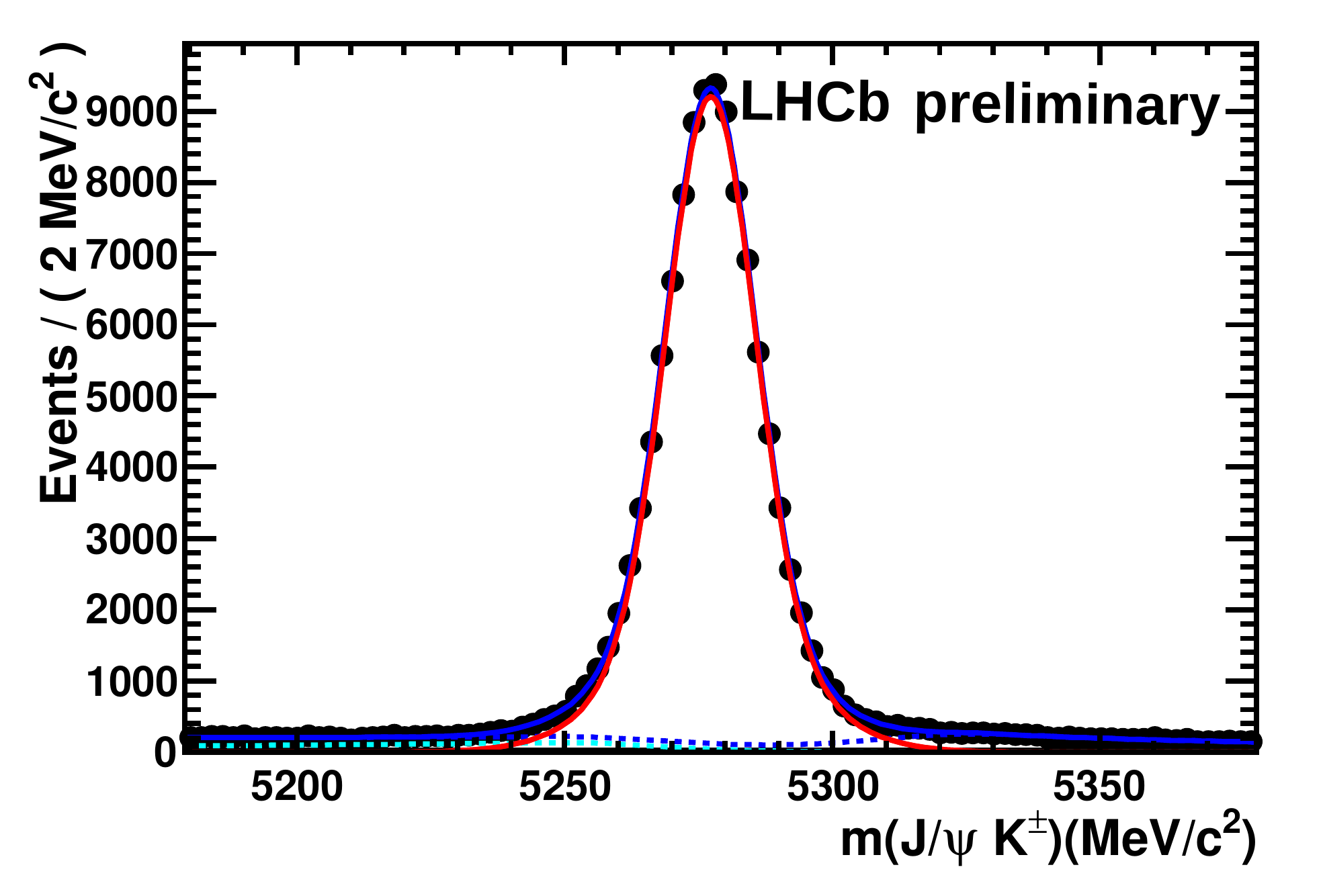}
 \includegraphics*[width=0.44\textwidth]{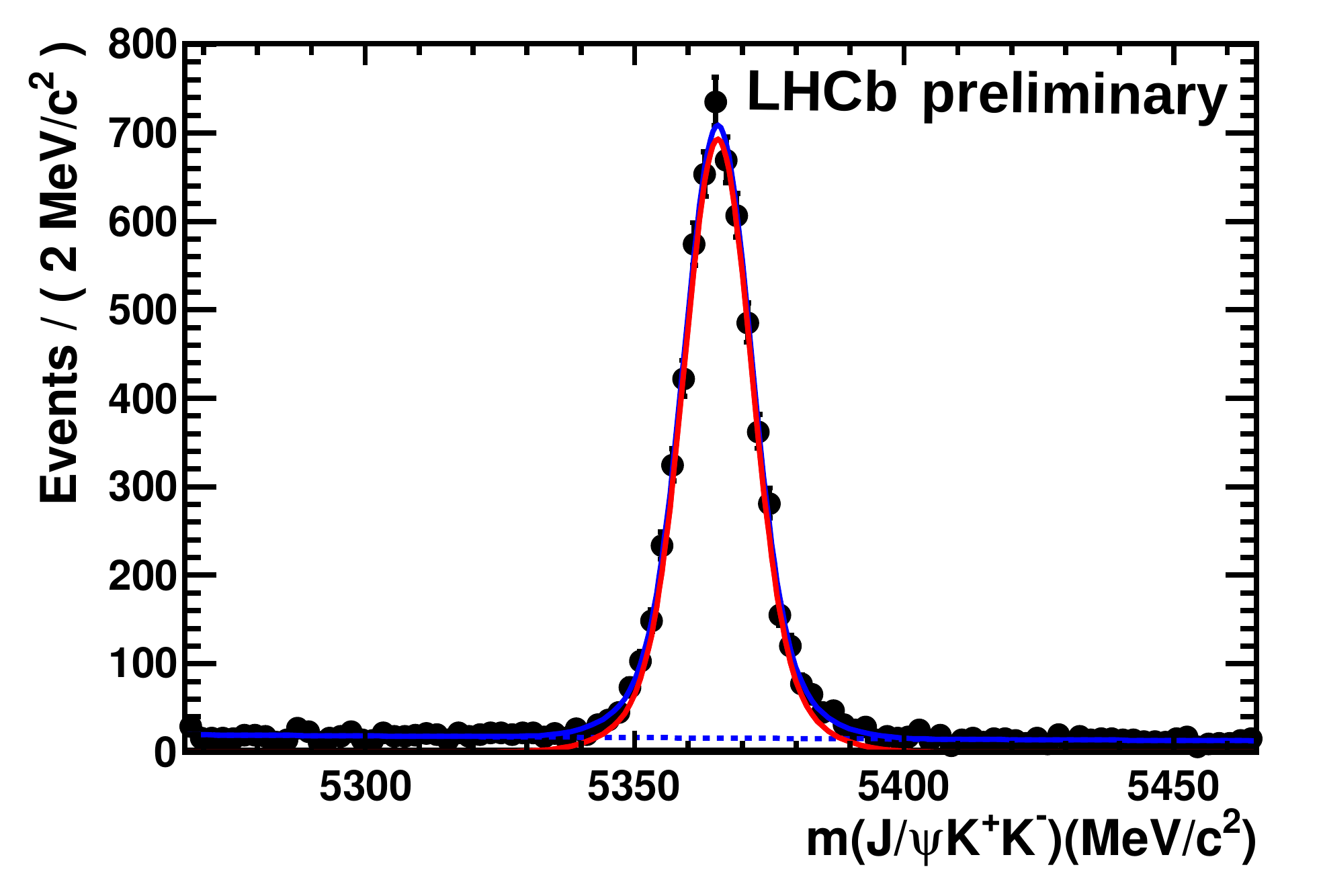}
\vspace{-4mm}
  \caption
  {Invariant mass distribution of the \BuJpsiK (left) and \BsJpsiPhi (right) candidates used in the normalization procedure.}  
  \label{fig:num_Bu}
\end{center}
\end{figure}

The number of expected signal events is evaluated by normalizing with three complementary channels of known BF, $B^+\to \Jpsi K^+$, $B^0_s\to \Jpsi \phi$ and $B^0 \to K^+\pi^-$.
The first two decays have similar trigger and muon identification efficiency to the signal but a different number of final-state particles, while the third channel has the same two-body topology but cannot be efficiently selected with the muon triggers.

The number of observed events $N_{\Bqmumu}$ is translated into a BF via the relation:
\begin{eqnarray}
\mathcal{B}  &=& {\mathcal{B}_{\rm norm}}\times\frac{\rm
\epsilon_{norm}^{REC}
\epsilon_{norm}^{SEL|REC}
\epsilon_{norm}^{TRIG|SEL}
}{\rm
\epsilon_{sig}^{REC}
\epsilon_{sig}^{SEL|REC}
\epsilon_{sig}^{TRIG|SEL}
}\times\frac{f_{\rm norm}}{f_{\Bq}}
\times\frac{N_{\Bqmumu}}{N_{\rm norm}} = \alpha_{\rm norm} \times N_{\Bqmumu}\,, \nonumber
\label{eq:normAlpha}
\end{eqnarray}
where $f_{\Bq}$ and $f_{\rm norm}$ are the probabilities that a $b$-quark fragments into a $\Bq$ and into the $b$-hadron relevant for the chosen calibration mode. 
We use LHCb measurement for $f_d/f_s = 3.745 \pm 0.295$~\cite{fdfs}.
${\cal B}_{\rm norm}$ is the branching fraction and $N_{\rm norm}$ the number of selected events of the normalization channel. 
The efficiency is separated into three factors: $\epsilon^{\rm REC}$ is the efficiency to reconstruct all the tracks of the decay including the geometrical acceptance of the
detector; $\epsilon^{\rm SEL|REC}$ is the efficiency to select the events which have been reconstructed; $\epsilon^{\rm TRIG|SEL}$ is the efficiency of the trigger on reconstructed and selected events. 
The ratios of reconstruction and selection efficiencies are estimated from simulations and checked with data, while the ratios of trigger efficiencies on selected events are determined from data~\cite{TisTos}.
Finally, $\alpha_{\rm norm}$ is the normalization factor.

Each $N_{\rm norm}$ is obtained with a fit to the invariant mass distribution of the normalization channel.
The invariant mass distributions for \BuJpsiK and \BsJpsiPhi are shown in Fig.~\ref{fig:num_Bu}, while the \BdToKpi yield is obtained via the \Bhh full fit as pictured in Fig.~\ref{fig:b2hh_fitinbins} top left.
The normalization factors calculated using the three channels give compatible results; the final normalization factors are weighted averages and read $\alpha_{B^0_{(s)} \to \mu\mu}= (9.84 \pm 0.91) \times 10^{-10}$ and $\alpha_{B^0 \to \mu\mu}= (2.59 \pm 0.15) \times 10^{-10}$.

\section{Results}
\label{sec:results}

\begin{figure}[tb]
\begin{center}
\includegraphics*[width=0.81\textwidth]{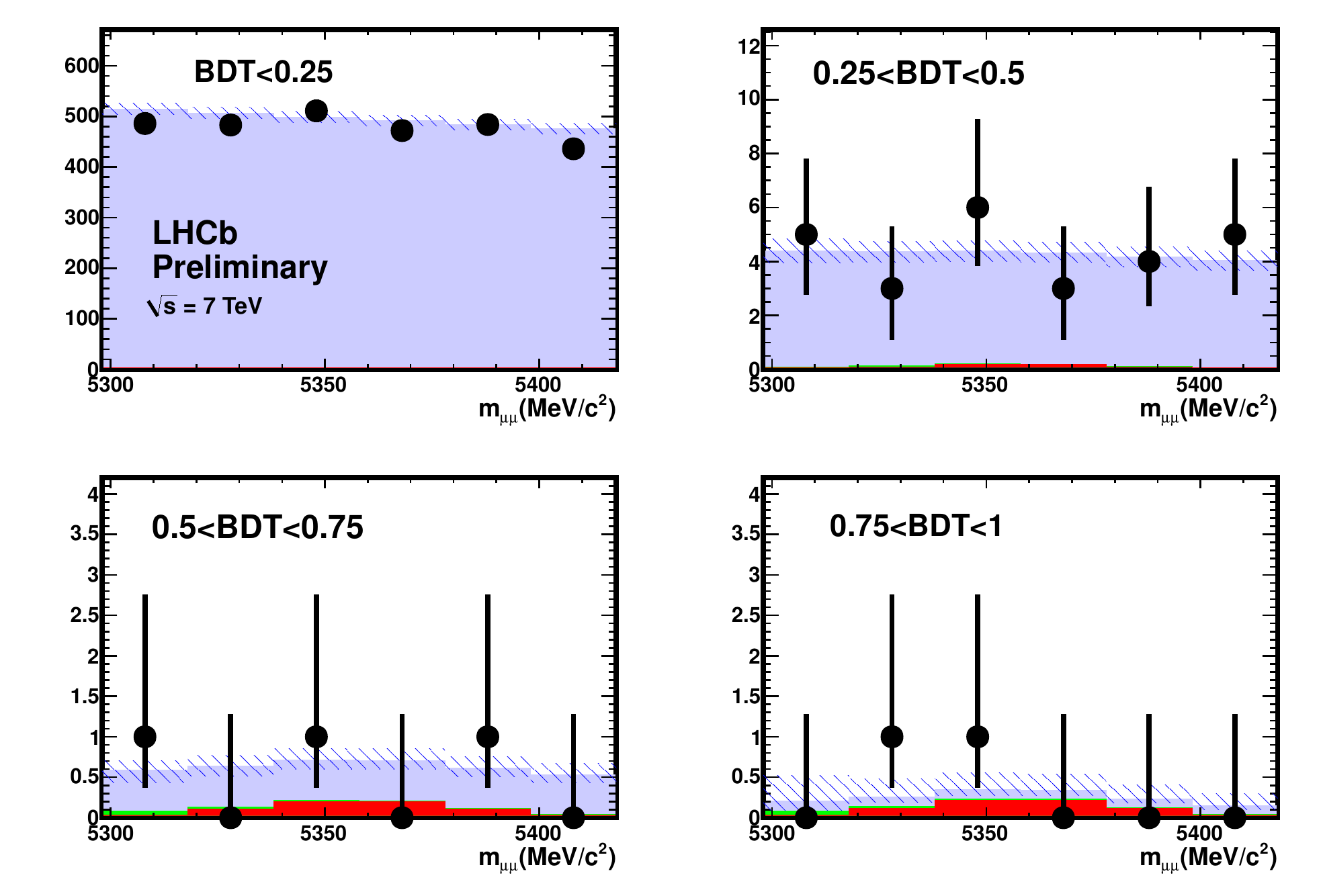}
\end{center}
\vspace{-5mm}
\caption{\Bsmumu: distribution of selected dimuon events in the invariant mass plane for the four BDT bins. 
The black dots are data, the light blue histogram shows the contribution of the combinatorial background, the green histogram shows the contribution of the \Bhh background and the red filled histogram the contribution of \Bsmumu signal events according to the SM rate.
The uncertainty on the expectations is shown by the hatched area. 
The uncertainty on  data in the first BDT bin is smaller than the size of the dots.}
\label{fig:fondo_bs}
\end{figure}

\begin{figure}[tb]
\begin{center}
\includegraphics*[width=0.81\textwidth]{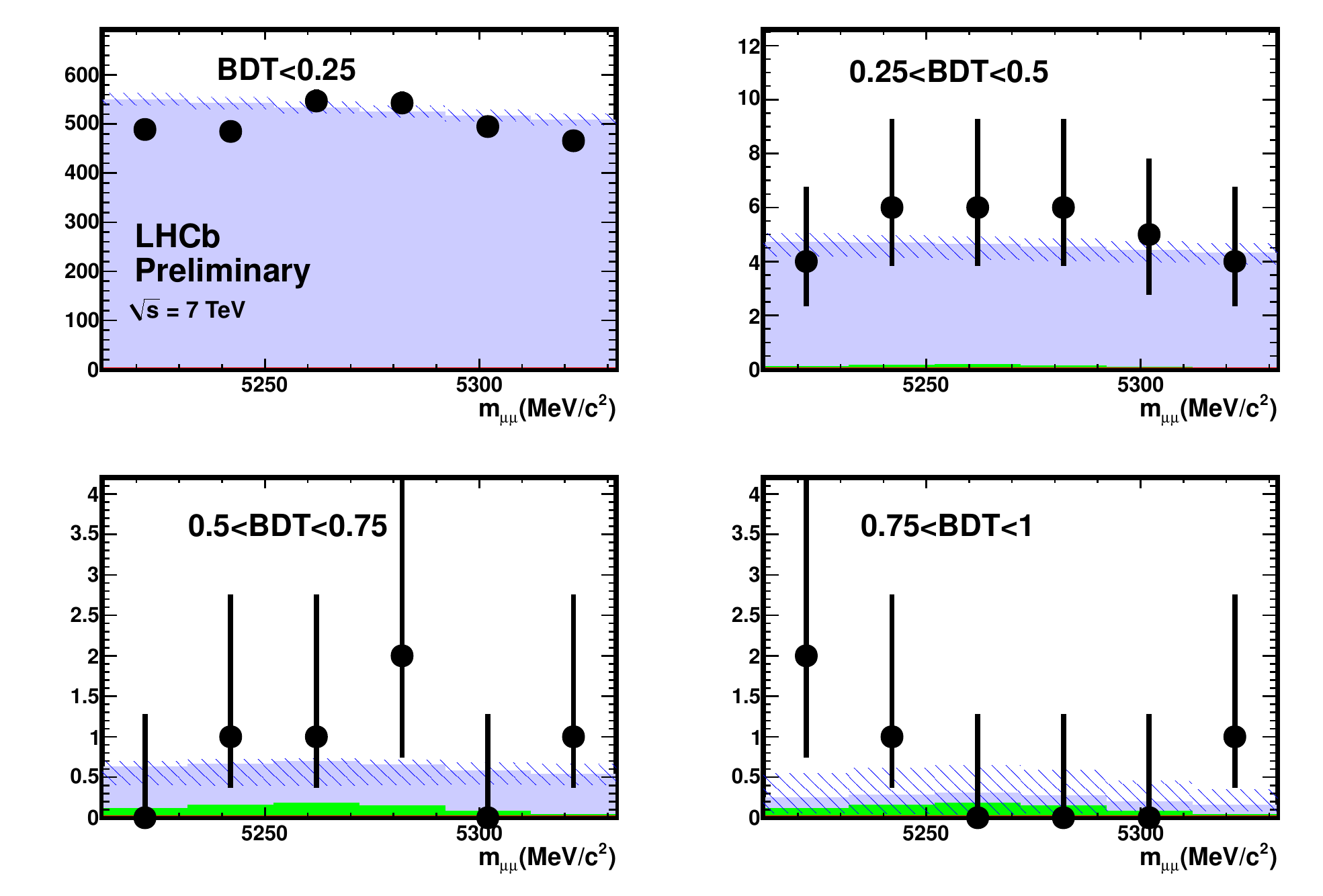}
\end{center}
\vspace{-5mm}
\caption{\Bdmumu: distribution of selected dimuon events in the invariant mass plane for the four BDT bins. 
The black dots are data, the light blue histogram shows the contribution of the combinatorial background and the green histogram shows the contribution of the \Bhh background.
The uncertainty on the expectations is shown by the hatched area.  
The uncertainty on data in the first BDT bin is smaller than the size of the dots.}
\label{fig:fondo_bd}
\end{figure}

The expected number of combinatorial background events,
the number of peaking background events as well as the number of expected signal events assuming a given BF hypothesis are computed for each 24 bins.
The distribution of the invariant mass in the four BDT bins is shown in Fig.~\ref{fig:fondo_bs} and~\ref{fig:fondo_bd} for \Bsmumu candidates and \Bdmumu candidates, respectively. The observed events are in good agreement with background expectations.
For \Bsmumu the observed events are also in good agreement with the background expectations and the presence of \Bsmumu events according to SM predictions.

The compatibility of the distribution of events in the invariant mass {\it vs} BDT plane with a given branching fraction hypothesis is evaluated using the \CLs method~\cite{Read_02}. 
This provides two estimators: \CLs is a measure of the compatibility of the observed distribution with the signal hypothesis, while \CLb is a measure of the compatibility with the background-only hypothesis. 

\begin{figure}[tbp]
\includegraphics[width=0.495\textwidth]{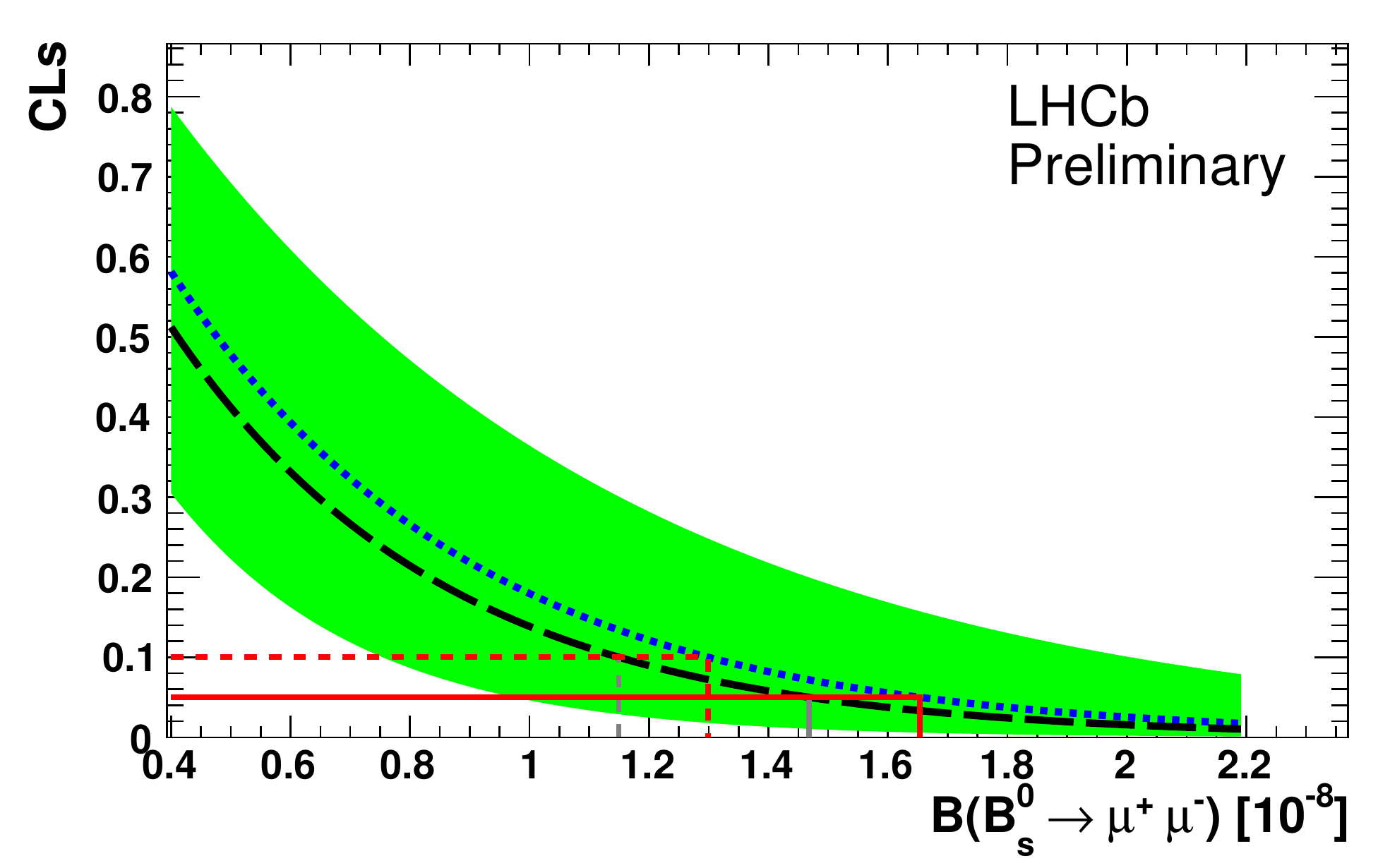}
\includegraphics[width=0.495\textwidth]{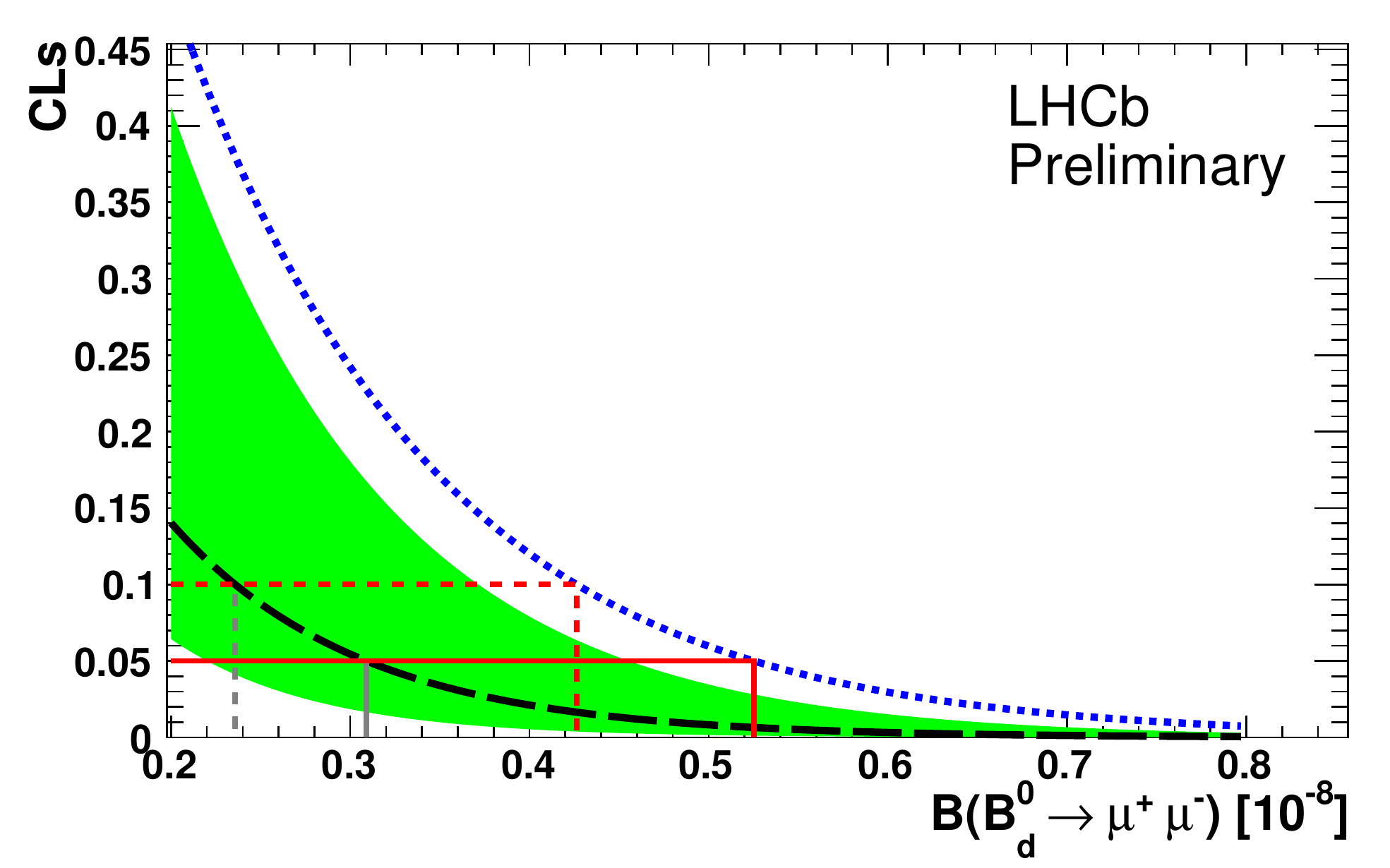}
\vspace{-4mm}
\caption
{Expected distributions of possible values of \CLs (dashed black line) under the hypothesis to observe a combination of background and signal events according to the SM rate in the case of \Bsmumu (left) and under the hypothesis to observe background only in the case of \Bdmumu  (right).
On both graphs, the green areas cover the region of $\pm 1 \sigma$ of compatible observations.
The observed distributions of \CLs as a function of the assumed branching fraction are shown as dotted blue lines for  \Bsmumu (left) and  \Bdmumu (right).
The measured upper limits at 90\% and 95\% C.L.\ are also shown.}
\label{fig:CLsBsmm_bkg}
\end{figure}

For the \Bsmumu decay, the distributions of expected \CLs values are shown as dashed (black) lines in Fig.~\ref{fig:CLsBsmm_bkg} (left) under the hypothesis to observe background plus SM events.
The green shaded areas cover the region of $\pm 1 \sigma$ of compatible observations.
The observed  \CLs as a function of the assumed branching fraction is shown as dotted (blue) line.
For the \Bdmumu decay, the distribution of possible values of \CLs is shown as dashed (black) line in Fig.~\ref{fig:CLsBsmm_bkg} (right) under the hypothesis to observe background events only.
The observed \CLs as a function of the assumed branching fraction is shown as a dotted (blue) line.


The upper limits read:
\begin{align}
\BRof\Bsmumu &< 1.3~(1.6)\times10^{-8}~{\rm at}~90\,\% ~(95\,\%)~{\rm C.L.,}  \nonumber \\
\BRof\Bmumu  &< 4.2~(5.2)\times10^{-9}~{\rm at}~90\,\% ~(95\,\%)~{\rm C.L.,} \nonumber
\end{align} 
while the expected values of the limits are $\BRof\Bsmumu < 1.2~(1.5)\times10^{-8}$ under the hypothesis to observe background plus SM events and $\BRof\Bmumu < 2.4~(3.1) \times 10^{-9}~{\rm at}~90\,\% ~(95\,\%)$ C.L.\ under the hypothesis to observe background events only.

\FloatBarrier

The comparison of the observed distribution of events with the expected background distribution  results in a  p-value (1-\CLb) of 20\,\% (21\,\%) for the \Bsmumu (\Bdmumu) decays.
In the case of \Bdmumu, the slightly low p-value is due to an excess of the observed 
events in the most sensitive BDT bin with respect to the background expectations. 
A larger data sample will allow to clarify the situation. 
In the case of  \Bsmumu, when a signal is included at the level expected in the Standard Model the p-value increases to 50\,\%.

A limit on \BRof\Bsmumu  has been extracted after combining the analysis on 300\invpb of 2011 data with the published analysis~\cite{LHCb_paper} on 37\invpb of 2010 data; the upper limits improve to:
\begin{eqnarray}
\BRof{\Bsmumu} (2010+2011) &<& 1.2 (1.5) \times 10^{-8}~{\rm at}~90\,\% (95\,\%)~{\rm C.L.}  \nonumber
\end{eqnarray} 
The above 95\% \CL upper limits are still about 4.7 times above the SM predictions for the \Bs and 51 times for the \Bd.

\bigskip 


\end{document}